    \documentstyle[nato,psfig,namedreferences]{crckapb}

\newcommand{\etal}{et~al.~}
\newcommand{\Msun}{\ifmmode{M_\odot}\else$M_\odot$~\fi}

\def\la{\mathrel{\mathchoice {\vcenter{\offinterlineskip\halign{\hfil
$\displaystyle##$\hfil\cr<\cr\noalign{\vskip1.5pt}\sim\cr}}}
{\vcenter{\offinterlineskip\halign{\hfil$\textstyle##$\hfil\cr<\cr
\noalign{\vskip1.0pt}\sim\cr}}}
{\vcenter{\offinterlineskip\halign{\hfil$\scriptstyle##$\hfil\cr<\cr
\noalign{\vskip0.5pt}\sim\cr}}}
{\vcenter{\offinterlineskip\halign{\hfil$\scriptscriptstyle##$\hfil
\cr<\cr\noalign{\vskip0.5pt}\sim\cr}}}}}

\begin{opening}
\title{FORMATION OF DISK GALAXIES: ON THE ANGULAR \protect\\
MOMENTUM PROBLEM, THE TULLY-FISHER \protect\\
RELATION AND MAGNETOHYDRODYNAMICS}

\author{JESPER SOMMER-LARSEN}
\institute{Theoretical Astrophysics Center\\
	   Juliane Maries Vej\\
	   DK-2100  Copenhagen {\O}, Denmark}

\end{opening}

\runningtitle{FORMATION OF DISK GALAXIES}

\begin{document}


\begin{abstract}
Two ways of possibly solving the angular momentum 
problem plaguing cold dark matter (CDM) {\it ab initio} simulations of disk 
galaxy formation are discussed: 1) Stellar
feedback processes and 2) Warm dark matter (WDM) rather than CDM.

In relation to the chemical evolution of disk galaxies
our simulations indicate that in case 1) the first generation of 
{\it disk} stars formed in disk galaxies like the
Milky Way should have an abundance about two dex below solar, in fairly 
good agreement with the lowest observed abundance of the metal-weak tail of
the Galactic thick disk. For the second case no such statements can be made
without further assumptions about the star-formation history of the 
galaxies.

We find that the $I$-band Tully-Fisher relation can be matched by 
WDM disk galaxy formation simulations provided $(M/L_I) \sim$ 0.8 for disk
galaxies, which Sommer-Larsen \& Dolgov (1999) argue is a reasonable 
value.

Finally it is discussed how the magnetic field strengths observed in galactic
disks can be obtained through disk galaxy {\it formation}, as an alternative
to the conventional dynamo hypothesis.

\end{abstract}
 
\section{Introduction}
\label{s:intro}

The formation of galactic disks is one of the most important unsolved
problems in astrophysics today. In the currently favored hierarchical
clustering framework, disks form in the potential wells of dark matter
halos as the baryonic material cools and collapses dissipatively. Fall \& 
Efstathiou (1980) have shown that
disks formed in this way can be expected to possess the observed
amount of angular momentum (and therefore the observed spatial extent
for a given mass and profile shape), but only under the condition that the
infalling gas retain most of its original angular momentum.

However, numerical simulations of this collapse scenario in the cold dark 
matter (CDM) cosmological context 
(e.g., 
Navarro \& Benz 1991,
Navarro \& White 1994,
Navarro, Frenk, \& White 1995)
have so far consistently indicated that when only cooling processes are
included the infalling gas loses too much angular momentum (by over
an order of magnitude) and the resulting disks are accordingly much
smaller than required by the observations.
This discrepancy is known as the {\em angular momentum problem} of
disk galaxy formation.
It arises from the combination of the following two facts:
a) In the CDM scenario the magnitude of linear density fluctuations  
$\sigma(M) = \langle(\delta M/M)^2\rangle^{1/2}$ increases steadily with 
decreasing
mass scale $M$ leading to the formation of non-linear, virialized structures
at increasingly early epochs with decreasing mass i.e.~the hierarchical
``bottom-up'' scenario.  b) Gas cooling
is very efficient at early times due to gas densities being generally
higher at high redshift as well as the rate of inverse Compton cooling also 
increasing very rapidly with redshift.  a) and b)
together lead to rapid condensation of small, dense gas clouds,
which subsequently lose energy and (orbital) angular momentum by dynamical
friction against the surrounding dark matter halo before they
eventually merge to form the central disk.
A mechanism is therefore needed that prevents, or at least delays,
the collapse of protogalactic gas clouds and allows the gas to
preserve a larger fraction of its angular momentum as it settles into
the disk. Two such possible solutions are discussed in section 2. 

In section 3 we present some new results from our WDM disk galaxy formation
simulations on the Tully-Fisher relation
and in section 4 we discuss how the magnetic field strengths of a few $\mu$G
observed in galactic disks can be obtained
via disk galaxy {\it formation}, as an alternative to disk dynamo 
amplification.

\section{Towards solving the angular momentum problem}

 Two ways of possibly solving the angular momentum problem have recently been
 discussed in the literature: a) by invoking the effects of stellar
 feedback processes from either single, more or less uniformly distributed
 stars or star-bursts and b) by assuming that the dark matter is ``warm''
 rather than cold. Both options lead to the suppression of the formation
 of early, small and dense gas clouds, for a) because the small gas clouds
 may be disrupted due to the energetic feedback of primarily type II
 super-nova explosions and for b) simply because fewer of the small and
 dense gas clouds form in the first place for WDM free-streaming masses
 $M_{f,{\rm WDM}} \sim 10^{10}$-$10^{11} \Msun$. 

\subsection{Stellar feedback processes}

\begin{figure}
\psfig{file=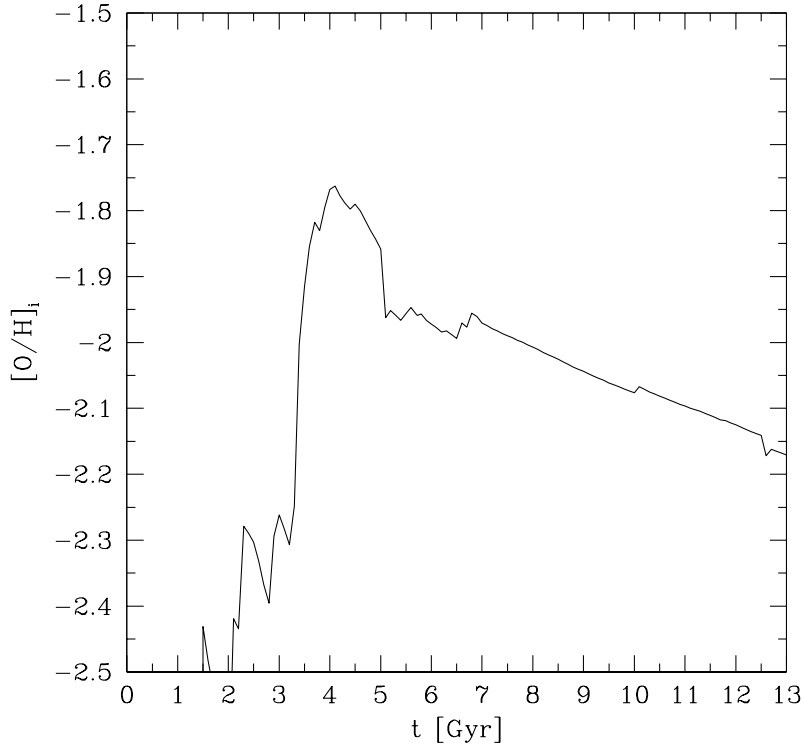,height=12cm,width=12cm}
\caption[]{The initial disk oxygen abundance resulting from infall of a 
mixture of enriched and unenriched gas onto the disk of a forming, Milky Way 
sized model galaxy. This abundance should be representative of the oxygen 
abundance of the first generation of {\it disk} stars formed and is fairly
consistent with the lowest found observationally for the metal-weak tail of
the Galactic thick disk.}
\end{figure}

Sommer-Larsen \etal (1999)
 showed that the feedback caused by a putative, early epoch of more or
 less uniformly distributed population III star formation was not
 sufficient to solve the angular momentum problem. Based on test 
 simulations they showed, however, that effects of feedback from star-bursts in
 small and dense protogalactic clouds might do that. Preliminary results of
 more sophisticated simulations incorporating stellar feedback processes
 in detail indicate that this is at least partly the case. Considerable
 fine-tuning seems to be required, however: About 2-3\% of the gas in
 the proto-galactic region of a forming disk galaxy should be turned into 
stars. If less
 stars are formed the feedback is not strong enough to cure the angular
 momentum problem and, vice versa, if more stars are formed during this
 fairly early phase of star-formation, the energetic feedback causes the
 formation of the main disks and thereby the bulk of the stars to be delayed 
too much compared to the observed star-formation history of the Universe.

This requirement of fine-tuning is advantageous, however, in relation to the 
early chemical evolution of disk galaxies, since the early star-formation
histories of the galaxies are then well constrained. Furthermore, as it 
is possible to track the elements produced and ejected by (primarily) type II
supernovae in the star-bursts one can determine the fraction of these
elements, which ultimately settle on the forming disk and hence
determine the rate and metallicity of the gas falling onto the disk.
In Figure 1 we show the time evolution of the oxygen abundance in a forming
disk as a result of infall of a mixture of enriched and unenriched gas
(neglecting the contribution of ejecta from stars formed subsequently in 
the disk). We have assumed a Salpeter IMF with $M_{low}=0.1 \Msun$ and
$M_{up}=60 \Msun$ and that a typical type II supernova ejects $\sim 2 \Msun$
of oxygen. This abundance can be regarded as the initial
abundance of the disk, its value depending on when star-formation 
subsequently commenced in the disk (note that such two-epoch star-formation
models have been advocated by, e.g., Chiappini, Matteucci \& Gratton 
1997).
As can be seen from the figure this initial disk abundance is of the
order $[O/H] \sim -2$. This is similar to the lowest abundance of the 
low-metallicity
tail of the Galactic thick disk -- see Beers \& Sommer-Larsen (1995).

\begin{figure}
\psfig{file=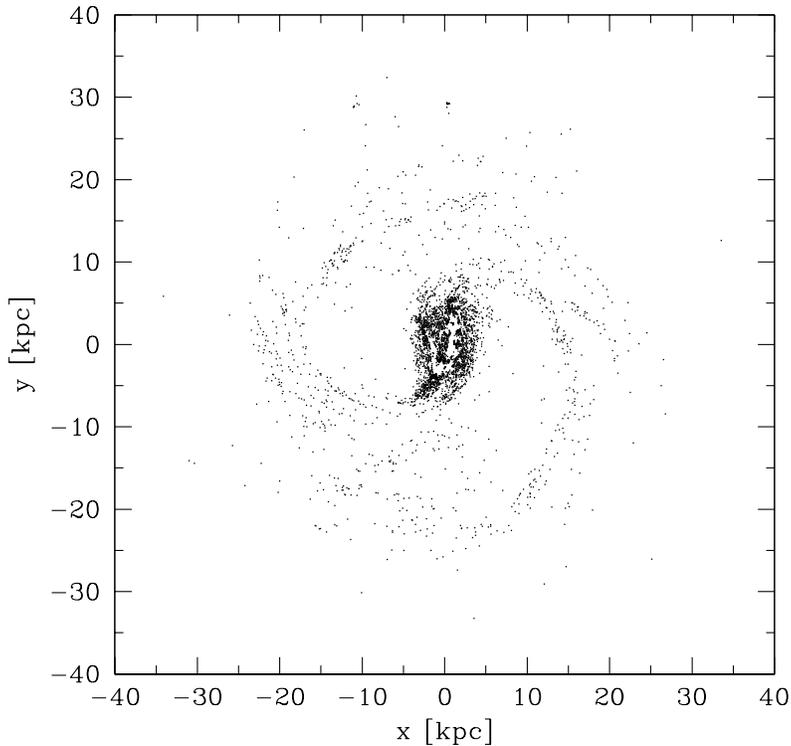,height=12cm,width=12cm}
\caption[]{Face-on view of a disk galaxy with characteristic circular velocity
$V_c \sim$ 300 km/s formed in a warm dark matter simulation with no conversion
of gas into stars (so the disk is purely gaseous). The mass of the disk is
$M_{disk} \sim 2\cdot 10^{11} \Msun$ and its specific angular momentum is
$j_{disk} \sim$ 2000 kpc km/s.}
\end{figure}

\subsection{Warm dark matter}

Another, more radical way of solving the angular momentum problem is to
abandon CDM altogether and assume instead that dark matter is ``warm''. Such
a rather dramatic measure not only proves very helpful in this respect, as 
will be discussed below, but may also be additionally motivated:
Recently, various possible shortcomings of the CDM cosmological scenario 
in relation to structure formation on galactic scales have been 
discussed
in the literature: 1) CDM possibly leads to the formation of too many
small galaxies relative to what is observed, i.e. the {\em missing satellites
problem} (e.g., Klypin \etal 1999).  2) Even if galactic winds due
to star-bursts can significantly reduce the number of visible dwarf
galaxies formed, sufficiently many of the small and tightly bound dark matter 
systems left behind can still survive to the present day in the dark matter
halos of larger galaxies like the Milky Way to possibly destroy the large,
central disks via gravitational heating, as discussed by Moore \etal 
(1999a).  3) The dark matter halos produced in CDM cosmological
simulations tend to have central cusps with $\rho_{DM}(r) \propto r^{-N},
N \sim 1-2$ (Dubinski \& Carlberg 1991, Navarro \etal 1996,
Fukushige \& Makino 1997, Moore \etal 1998, Kravtsov \etal 
1998, Gelato \& 
Sommer-Larsen 1999). This is in disagreement with the flat, central
dark matter density profiles (cores) inferred from observations of the 
kinematics of dwarf and low surface brightness galaxies (e.g., Burkert 
1995, de Blok \& McGaugh 1997,
Kravtsov \etal 1998, Moore \etal 1999b, but see also 
van den Bosch \etal 1999).

The first two problems may possibly be overcome by invoking warm dark
matter (WDM) instead of CDM: On mass scales less than the free-streaming mass,
$M \la M_{f,{\rm WDM}}$,
the growth of the initial density fluctuations in the Universe is suppressed
relative to CDM due to relativistic free-streaming of the warm dark matter
particles. In conventional WDM theory these become non-relativistic at
redshifts $z_{nr} \sim 10^6$-$10^7$ for $m_{{\rm WDM}} \sim$ 1 keV, which is the
characteristic WDM particle mass required to give sub-galactic to galactic 
free-streaming masses.
As a consequence of this suppression, fewer low mass galaxies (or 
``satellites'') are formed
cf., e.g., Moore \etal (1999a) and Sommer-Larsen \& Dolgov 
(1999, SD99). The 
central cusps problem may be more generic (Huss \etal 1999 and
Moore \etal 1999b),
but WDM deserves further attention also on this point.

SD99 show that the angular momentum problem
may be resolved by going from cold to warm dark matter,
with characteristic free-streaming mass $M_{f,{\rm WDM}} \sim 
10^{10}$-$10^{11} \Msun$, and without having to invoke
effects of stellar feedback processes at all.
The reason why this kind of warm dark matter leads to a solution of the
angular momentum problem is that because of the suppression of density 
fluctuations on sub-galactic scales relative to CDM the formation of a
disk galaxy becomes a much more coherent and gentle process enabling the
infalling, disk-forming gas to retain much more of its original angular
momentum. In fact SD99 find it 
likely that the angular momentum problem can be {\it completely} resolved
by going to the WDM structure formation scenario, which is more than can
be said for the CDM+feedback approach so far. In Figure 2 we show a
face-on view of a disk galaxy with characteristic circular velocity (where the
rotation curve is approximately constant) $V_c \sim$
300 km/s formed in a WDM simulation (in this simulation
gas was not converted into stars). Clearly it is no
longer a problem to form extended, high angular momentum disks in fully
cosmological simulations. In comparison the extent of typical disks formed
in ``passive'' CDM simulations (i.e. simulations not incorporating the effects
of stellar feedback processes) is less than 1 kpc -- see, e.g., Sommer-Larsen
\etal (1999).

Unlike the CDM+feedback solution, one does not get a constraint
on the early star-formation histories of the proto-galaxies, so no statements 
about the abundance of the first generation of disk stars can be made without 
further assumptions.

SD99 discuss possible physical candidates for WDM 
particles and find that
the most promising are neutrinos with weaker or stronger interactions than
normal, majorons (light pseudogoldstone bosons), or mirror or shadow world
neutrinos.

\begin{figure}
\psfig{file=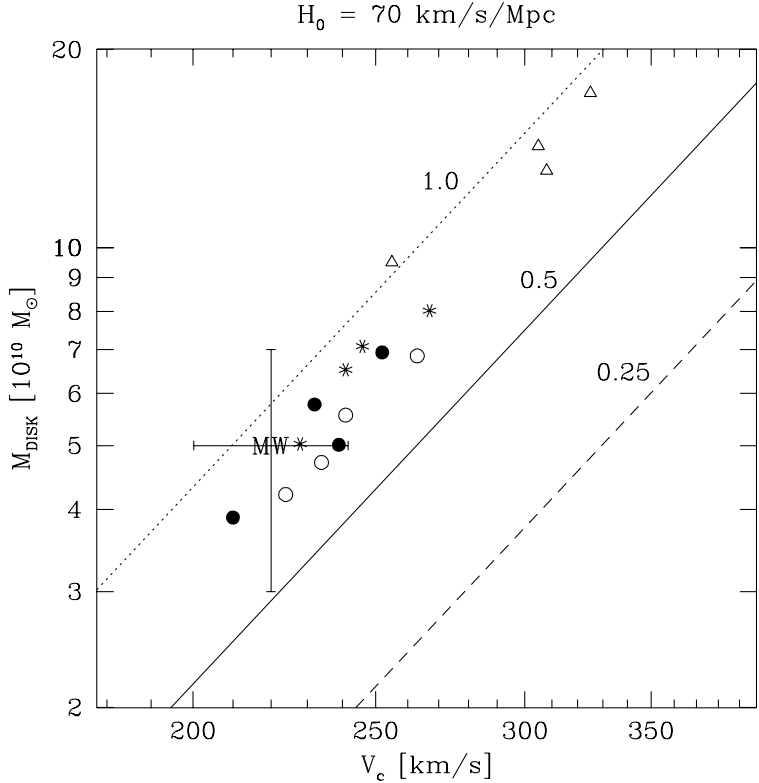,height=12cm,width=12cm}
\caption[]{The mass vs. characteristic circular velocity ``Tully-Fisher'' 
relation for the final disks formed in 16 WDM simulations of Sommer-Larsen 
\& Dolgov (for $H_0$=70 km/s/Mpc).
Also shown is the observed $I$-band TF relation of
Giovanelli \etal converted to mass assuming ($M/L_I$) = 0.25 ({\it dashed
line}), 0.5 ({\it solid line}) and 1.0 ({\it dotted line}) and $H_0$=70 
km/s/Mpc. Finally, the
symbol ``MW'' with errorbars shows the likely range of the total, baryonic mass
and characteristic circular velocity of the Milky Way.}
\end{figure}

\section{The Tully-Fisher relation}
In Figure 3 we show the cooled-out disk mass $M_{disk}$ at redshift $z$=0 as
a function of the characteristic circular velocity $V_c$ of model galaxies
formed in our WDM simulations (assuming a Hubble parameter $H_0$=70 km/s/Mpc).
Also shown is the $I$-band Tully-Fisher (TF)
relation of Giovanelli \etal (1997) converted to mass assuming $I$-band
mass-to-light ratios $(M/L_I)$=0.25, 0.5 and 1.0 in solar units and 
$H_0$=70 km/s/Mpc. Finally, the baryonic mass of the
Milky Way, estimated in a completely independent way, is shown (see SD99 for 
details). As can be seen from the figure we can match
the slope of the TF relation very well assuming a constant $(M/L_I)$. To get
the normalization right a $(M/L_I) \sim 0.8$ is required. SD99 argue that this
is quite a reasonable value in comparison
with various dynamical and spectrophotometric estimates. Moreover, it is 
clearly gratifying that the Milky Way data point falls right on top of the
theoretical as well as observational $M_{disk}$-$V_c$ relations (for
$(M/L_I) \sim 0.8$, $H_0$=70 km/s/kpc).

Steinmetz \& Navarro (1999) and Navarro \& Steinmetz (1999)
find a discrepancy between the observed and ``theoretical'' TF on the basis
of CDM simulations of disk galaxy formation. It is hence possible that
WDM helps out also on this point, but this has to be checked with more
detailed simulations.

\begin{figure}
\psfig{file=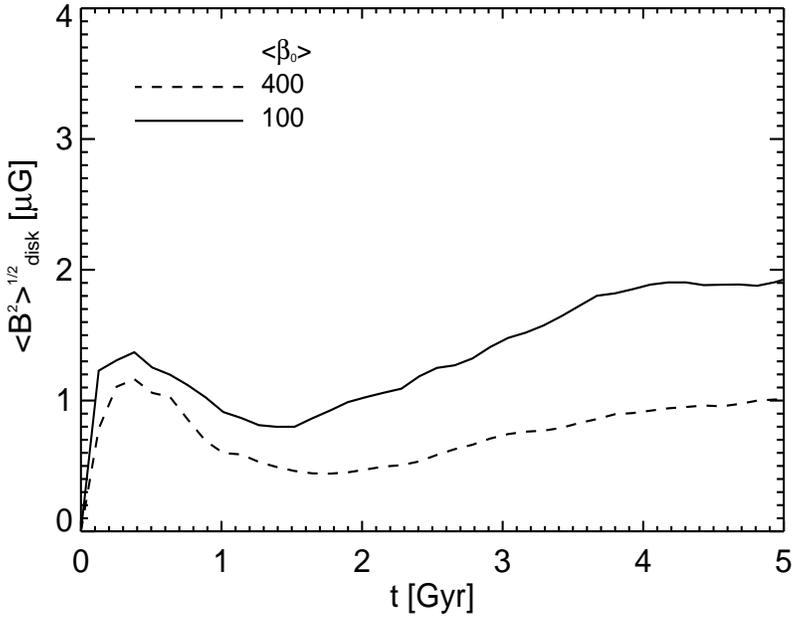,height=9cm,width=12cm}
\caption[]{The temporal evolution of the average field strength in the disk
for two values of $\langle\beta_0\rangle$.}
\end{figure}

\begin{figure}
\psfig{file=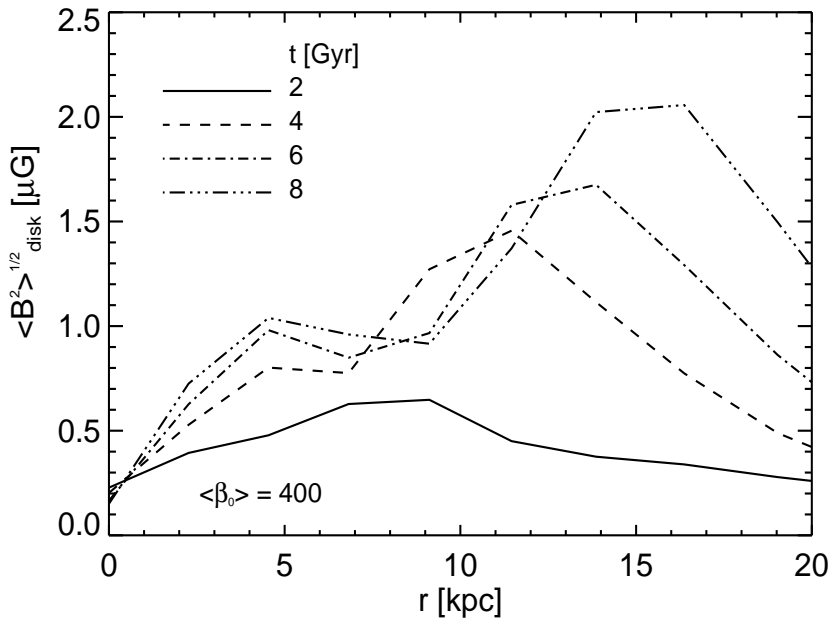,height=9cm,width=12cm}
\caption[]{The azimuthally averaged field strength in the disk at various
times for the $\langle\beta_0\rangle$=400 experiment.}
\end{figure}

\section{Magnetic fields in galactic disks and disk galaxy formation}

R{\"o}gnvaldsson (1999) showed how the typical magnetic field strengths
observed in galactic disks can be explained as a result of disk galaxy
{\it formation}, as an alternative to the usually assumed dynamo amplification
of an initially very weak magnetic field in the disk: Hot, virialized gas 
($T \sim 2\cdot 10^6$ K) in a dark matter halo is
assumed to initially follow the dark matter distribution and to be
rotating slowly, corresponding to a spin-parameter $\lambda \sim$ 0.05,
typical of galactic, dark matter halos. The hot gas is assumed to be threaded
by a weak and random magnetic field. As the hot gas cools 
radiatively, gravity forces it to flow inwards and due to the spin it forms
a growing, cold, galactic disk in the central parts of the dark matter halo. 
The
magnetic field follows the cooling gas inwards and is strongly amplified by
compression and shear in the forming disk. R{\"o}gnvaldsson (1999)             carried
out magnetohydrodynamical (MHD) simulations of this process using an eulerian
mesh MHD code.  The simulations were run with various 
initial magnetic field strengths
in the hot gas, parameterized by the initial ratio between the
gas pressure and magnetic pressure
$\langle\beta_0\rangle = \langle\frac{P_{gas}}{B^2/8\pi}\rangle_0$.
The temporal evolution of the average magnetic field strength in the disk
gas is shown in Figure 4. For weak initial fields 
($\langle\beta_0\rangle$=100-400 was taken as a starting point, since
these values are typical values in the hot, 
intergalactic gas in clusters of galaxies) the
average magnetic field strength grows gradually from about $t$=1 Gyr (after
an initial relaxation phase). The average values of 1-2$\mu$G reached after
about 5 Gyr are quite reasonable for typical disk galaxies, indicating a 
route to the explanation of the magnetic field strengths observed
in galactic disks alternative to the usual dynamo one.

Another aspect of the growth of the field strength is reflected in the radial
average in the disk, shown in Figure 5 at various times for 
a simulation with $\langle\beta_0\rangle$=400.
The field strength is always highest in the outermost part of the growing 
disk, since 
the fieldlines brought in with the cooling flow are stacked on top of the 
already existing field there and the field is further amplified by the disk 
shear.

\section{Acknowledgements}

I have
benefited from comments by {\"O}rn{\'o}lfur R{\"o}gnvaldsson,
Sasha Dolgov and Jens Schmalzing and
thank the organizers for a magnificent  
conference. This work was supported by Danmarks Grundforskningsfond through 
its support for the establishment of the Theoretical Astrophysics 
Center.


\begin{thebibliography}{}
\bibitem[1999]{} Beers, T. C., \& Sommer-Larsen, J. 1995, 
{\it Astrophys. J. Suppl.}, {\bf 96}, 175
\bibitem[1999]{} van den Bosch, F. C., \etal 1999, {\it Astron. J.}, submitted 
(astro-ph/9911372)
\bibitem[1995]{} Burkert, A. 1995, {\it Astrophys. J.}, {\bf 488}, L55
\bibitem[1997]{} de Blok, W. \ J. \ G., \& McGaugh, S. \ S. 1997, {\it MNRAS}, {\bf 290}, 533
\bibitem[1997]{} Chiappini, C., Matteucci, F, \& Gratton, R. 1997, 
{\it Astrophys. J.}, {\bf 477}, 765
\bibitem[1991]{} Dubinski, J., \& Carlberg, R. 1991, {\it Astrophys. J.}, {\bf 378}, 496
\bibitem[1980]{} Fall, S. M., \& Efstathiou, G. 1980, {\it MNRAS}, {\bf 193}, 189 
\bibitem[1997]{} Fukushige, T., \& Makino, J. 1997, {\it Astrophys. J.}, {\bf 477}, L9
\bibitem[1999]{} Gelato, S., \& Sommer-Larsen, J. 1999, {\it MNRAS}, {\bf 303}, 321
\bibitem[1997]{} Giovanelli, R., \etal 1999, {\it Astrophys. J.}, {\bf 477}, L1
\bibitem[1999]{} Huss, A., Jain, B., \& Steinmetz, M. 1999, {\it Astrophys. J.}, {\bf 517}, 64
\bibitem[1999]{} Klypin, A., Kravtsov, A. V., Valenzuela, O., \& Prada, F.
1999, {\it Astrophys. J.}, {\bf 523}, 32
\bibitem[1998]{} Kravtsov, A. \ V., Klypin, A. \ A., Bullock, J. \ S., \&
Primack, J. \ R. 1998, {\it Astrophys. J.}, {\bf 502}, 48
\bibitem[1998]{} Moore, B., Governato, F., Quinn, T., Stadel, J., \& Lake, G. 1998, {\it Astrophys. J.}, {\bf 499}, L5
\bibitem[1999a]{} Moore, B., Ghinga, S., Governato, F., Lake, G., Quinn, T., Stadel, J., \& Tozzi, P. 1999a, {\it Astrophys. J.}, {\bf 524}, L19
\bibitem[1999b]{} Moore, B., Quinn, T., Governato, F., Stadel, J., \&  Lake, G. 1999b, {\it MNRAS}, submitted (astro-ph/9903164)
\bibitem[1991]{} Navarro, J. F., \& Benz, W. 1991, {\it Astrophys. J.}, {\bf 380}, 320
\bibitem[1994]{} Navarro, J. F., \& White, S. D. M. 1994, {\it MNRAS}, {\bf 267}, 401
\bibitem[1995]{} Navarro, J. F., Frenk, C. S., \& White, S. D. M. 1995, {\it MNRAS}, {\bf 275}, 56
\bibitem[1996]{} Navarro, J. F., Frenk, C. S., \& White, S. D. M. 1996, {\it MNRAS}, {\bf 462}, 563
\bibitem[1999]{} Navarro, J. F., \& Steinmetz, M. 1999, {\it Astrophys. J.}, in press (astro-ph/9908114)
\bibitem[1999]{} R{\"o}gnvaldsson, {\"O}. 1999, PhD thesis, University of
Copenhagen
\bibitem[1999]{} Sommer-Larsen, J., Gelato, S., \& Vedel. H.
1999, {\it Astrophys. J.}, {\bf 519}, 501 
\bibitem[1999]{} Sommer-Larsen, J., \& Dolgov, A. 1999, {\it Astrophys. J.}, submitted
(astro-ph/9912166, SD99)
\bibitem[1999]{} Steinmetz, M., \& Navarro, J. F. 1999, {\it Astrophys. J.}, {\bf 513}, 555
\end{thebibliography}
\end{document}